\documentclass[conference]{IEEEtran}
\IEEEoverridecommandlockouts
\usepackage{xspace}
\usepackage{caption}
\usepackage{subcaption}
\usepackage{cite}
\usepackage{amsmath,amssymb,amsfonts}
\usepackage{algorithmic}
\usepackage{graphicx}
\usepackage{textcomp}
\usepackage{xcolor}
\usepackage{xurl}
\usepackage{listings}
\usepackage{comment}
\usepackage[hidelinks]{hyperref}
\def\BibTeX{{\rm B\kern-.05em{\sc i\kern-.025em b}\kern-.08em
    T\kern-.1667em\lower.7ex\hbox{E}\kern-.125emX}}

\newcommand{\tool}{NMO\xspace}

\begin{document}

\title{Multi-level Memory-Centric Profiling on ARM Processors with ARM SPE \thanks{pre-print submitted for publication}
}

\author{\IEEEauthorblockN{Samuel Miksits}
\IEEEauthorblockA{\textit{KTH Royal Institute of Technology}\\
Stockholm, Sweden \\
smiksits@kth.se}
\and
\IEEEauthorblockN{Ruimin Shi}
\IEEEauthorblockA{\textit{KTH Royal Institute of Technology}\\
Stockholm, Sweden \\
ruimins@kth.se}
\and
\IEEEauthorblockN{Maya Gokhale}
\IEEEauthorblockA{\textit{Lawrence Livermore National Laboratory}\\
Livermore, USA \\
gokhale2@llnl.gov}
\and
\IEEEauthorblockN{Jacob Wahlgren}
\IEEEauthorblockA{\textit{KTH Royal Institute of Technology}\\
Stockholm, Sweden \\
jacobwah@kth.se}
\and
\IEEEauthorblockN{Gabin Schieffer}
\IEEEauthorblockA{\textit{KTH Royal Institute of Technology}\\
Stockholm, Sweden \\
gabins@kth.se}
\and
\IEEEauthorblockN{Ivy Peng}
\IEEEauthorblockA{\textit{KTH Royal Institute of Technology}\\
Stockholm, Sweden \\
ivybopeng@kth.se}
}

\maketitle

\begin{abstract}
High-end ARM processors are emerging in data centers and HPC systems, posing as a strong contender to x86 machines. Memory-centric profiling is an important approach for dissecting an application's bottlenecks on memory access and guiding optimizations. Many existing memory profiling tools leverage hardware performance counters and precise event sampling, such as Intel PEBS and AMD IBS, to achieve high accuracy and low overhead. In this work, we present a multi-level memory profiling tool for ARM processors, leveraging Statistical Profiling Extension (SPE). We evaluate the tool using both HPC and Cloud workloads on the ARM Ampere processor. Our results provide the first quantitative assessment of time overhead and sampling accuracy of ARM SPE for memory-centric profiling at different sampling periods and aux buffer sizes. 
\end{abstract}


\section{Introduction}
In the past decades, high-performance computing (HPC) platforms have been dominated by x86 processors, where the majority of supercomputers on the Top 500 list~\cite{top500} are based on Intel and AMD processors. Recently, high-end server-class ARM processors have started showing their potential as a replacement architecture for HPC systems and data centers. Exemplified by Fugaku, an A64FX-based Supercomputer now ranking fourth on the Top 500 list~\cite{top500}. Moreover, cloud providers like Amazon are now providing massive computing power via their AWS Graviton ARM Processor, and Nvidia's latest Grace Hopper superchip is based on ARM Neoverse V2~\cite{schieffer2024harnessing}. Both Graviton and Grace CPUs feature the ARM scalable vector extension (SVE) for boosting computing power at high energy efficiency. This trend of emerging ARM processors in HPC and data centers is expected to continue due to diminishing gains from x86 and increasing energy concerns~\cite{yokoyama2019survey, lopez2024genarchbench}.

Profiling tools are critical for dissecting performance bottlenecks and guiding developers to fully exploit the underlying hardware. On different processors, vendor-provided profiling tools, such as Intel's VTune, AMD's uProf, and ARM's MAP, have been widely used in identifying hotspot code regions. Complementary to code-region-based profiling, \textit{memory-centric profiling} provides a way to identify memory regions with bottlenecks during execution. For instance, several previous memory-centric profiling tools have been developed to identify hot memory regions that cause extensive false sharing, cache misses, or remote NUMA accesses. When a memory problem is revealed, different optimization techniques, such as padding, data placement, and memory interleaving, can be derived to optimize the performance. Such memory-centric profiling tools have become increasingly important as the memory hierarchy on modern processors is becoming deeper and more complex. 

Memory-centric profiling typically covers various metrics on memory access patterns and memory usage. These memory metrics can help application users select suitable memory resources or guide developers to derive effective optimization strategies. At the high level, a memory profiling tool would provide the basic utilization of the memory system. Memory footprint is one common indicator of an application's need for memory capacity. The emergence of High-bandwidth memory (HBM) combined with DDR DRAM in heterogeneous memory systems further demands a thorough understanding of an application's memory bandwidth utilization pattern, so that only bandwidth-sensitive applications need to be allocated with HBM resources. Finally, data locality, both spatial and temporal, could present vast differences in massive concurrent applications, which are commonly found in shared-memory parallel applications in HPC systems. Thus, a precise understanding of memory access patterns in multi-threaded applications, in addition to memory capacity and bandwidth usage patterns, become the three pillars of memory-centric profiling.  

Existing memory profiling tools are often based on one of the three mechanisms: \textit{binary instrumentation}, \textit{compiler analysis}, and \textit{hardware performance counters}. Among them, memory profiling tools that target runtime adaptation and thus need low overhead, are often based on hardware performance counters. Precise event sampling, an advanced feature supported on Performance Monitoring Units (PMUs) on Intel and AMD, is one low-overhead approach for capturing precise memory access patterns in multi-threaded applications~\cite{soramichi2017quantitative}. For instance, Intel processors typically support Processor Event-Based Sampling (PEBS) and AMD supports Instruction-Based Sampling (IBS), among which many memory profiling tools have been developed~\cite{liu2015scaanalyzer,boehme2016caliper,soramichi2017quantitative,sasongko2023precise}. Recently, ARM processors have gained similar support for precise event sampling in Armv8.2~\cite{armneoversev1} through the Statistical Profiling Extension (SPE). As more ARM processors are emerging on HPC systems, there is a clear need for memory profiling tools for memory-centric profiling on ARM processors. However, currently, there is still little quantitative evaluation of their accuracy, overhead, and limitations.

In this work, we design a memory-centric profiling tool called \tool for ARM processors. Besides memory bandwidth and capacity usage, \tool specifically leverages ARM's Statistical Profiling Extension (SPE) to enable memory-region-based profiling. We evaluate \tool on an ARM Ampere processor in five applications, including Stream, Rodinia's CFD and BFS, and CloudSuite's Page Rank and In-memory Analytics. Our results quantitatively evaluate the time overhead and sampling accuracy of ARM SPE at different sampling periods and memory buffer sizes. At 3000 and 4000 sampling periods, the ARM SPE profiling achieves the highest accuracy above $94\%$ at a time overhead of $0.2\%$--$3.3\%$. We found that a high sampling frequency, i.e., sampling periods lower than 2000, causes significant sample drops and low accuracy. Meanwhile, a memory buffer of 16--32~pages (64KB each on our ARM testbed) result in the optimal overhead and accuracy in test applications. 

In summary, we made the following contributions in this work.
\begin{itemize}
    \item We describe the design of a multi-level memory-centric profiling tool.
    \item We provide a prototype implementation called \tool using ARM PMU and Statistical Profiling Extension (SPE).
    \item We evaluate \tool in five HPC and Cloud benchmarks on an ARM Ampere processor.
    \item We provide the first quantitative evaluation of time overhead and sampling accuracy of ARM SPE for memory-centric profiling.
    \item Our results show that ARM SPE can achieve sampling accuracy above $94\%$ at a small time overhead between $0.2\%$ and $3.3\%$ in tested applications.
\end{itemize}
\section{Background}

\textbf{Server-class ARM Processors.} For decades, x86 processors have dominated high-end computing platforms such as HPC systems. As evident from the Top 500 list~\cite{top500}, most top-ranking supercomputers are based on Intel machines, with the recent emergence of AMD machines, such as Frontier and LUMI supercomputers positioning as the fastest HPC systems in the world and Europe, respectively. The IBM Power architecture is also used in top supercomputers such as the Summit and Sierra pre-exascale supercomputers. For a long time, ARM processors have dominated a different market, namely mobile devices, where energy efficiency rather than performance is the top priority. 

The first top500 ARM machine was Astra supercomputer, based on ARMv8.1 and NEON SIMD, at Sandia National Laboratory~\cite{astra}. The emergence of ARM architecture on high-end computing facilities starts with the ARMv8.x paired with the scalable vector extension (SVE)~\cite{stephens2017arm}. For instance, the Fukagu supercomputer now ranks No. 4 in the Top500 list and is based on the A64FX processor, an ARM architecture with SVE implemented by Fujitsu. Other major vendors like Nvidia and cloud providers like Amazon, have also based their server-class processors, including Grace CPU and AWS Graviton CPU, on the ARM architecture. In recent years, Europe has increasingly invested in ARM-based chip design for sovereign autonomy in the chip manufacturing sector, with the first European Exascale machine, Jupiter, to use ARM-based CPUs~\cite{jupiter}.  

\textbf{Memory Profiling} is an important approach for detecting memory access patterns, identifying hotspot memory regions, and deriving optimization strategies. The vast number of existing tools are mostly based on three fundamental mechanisms, namely \textit{binary instrumentation}, \textit{compiler analysis}, and \textit{hardware performance counters}. Profiling tools based on hardware performance counters incur much lower overhead compared to those based on binary instrumentation, as represented by the Intel Pin tools. 

While tools based on compiler analysis can leverage detailed information by analyzing memory access instructions, and derive code transformation to achieve automatic optimization, they have limitations in adapting to the changes in the underlying memory systems, e.g., changes in cache levels.  Also, they might fall short in complex memory access patterns via pointer aliasing. As the memory systems on HPC systems are getting deeper and more complicated, memory profiling should be able to reflect the impact of differences in memory hierarchy on computing platforms, as an application tuned for one may exhibit a sub-optimal memory access pattern on another. Therefore, many memory profiling tools, especially those targeting runtime adaptation, have been designed based on hardware performance counters. Due to the prevalence of x86 machines, they are mostly based on the processor-specific precise event sampling feature, namely Intel's PEBS and AMD's IBS.

On ARM processors, the Statistical Profiling Extension (SPE) provides similar precise event sampling features. It was first introduced to address limitations in the previous standard PMU in ARM, where limited memory information is provided. SPE is defined as part of Armv8-A architecture (starting from v8.2), which provides hardware-based statistical sampling.  

\subsection{ARM Statistical Profiling Extension}
 \begin{figure}[ht]
    \centering
    \includegraphics[width=\linewidth]{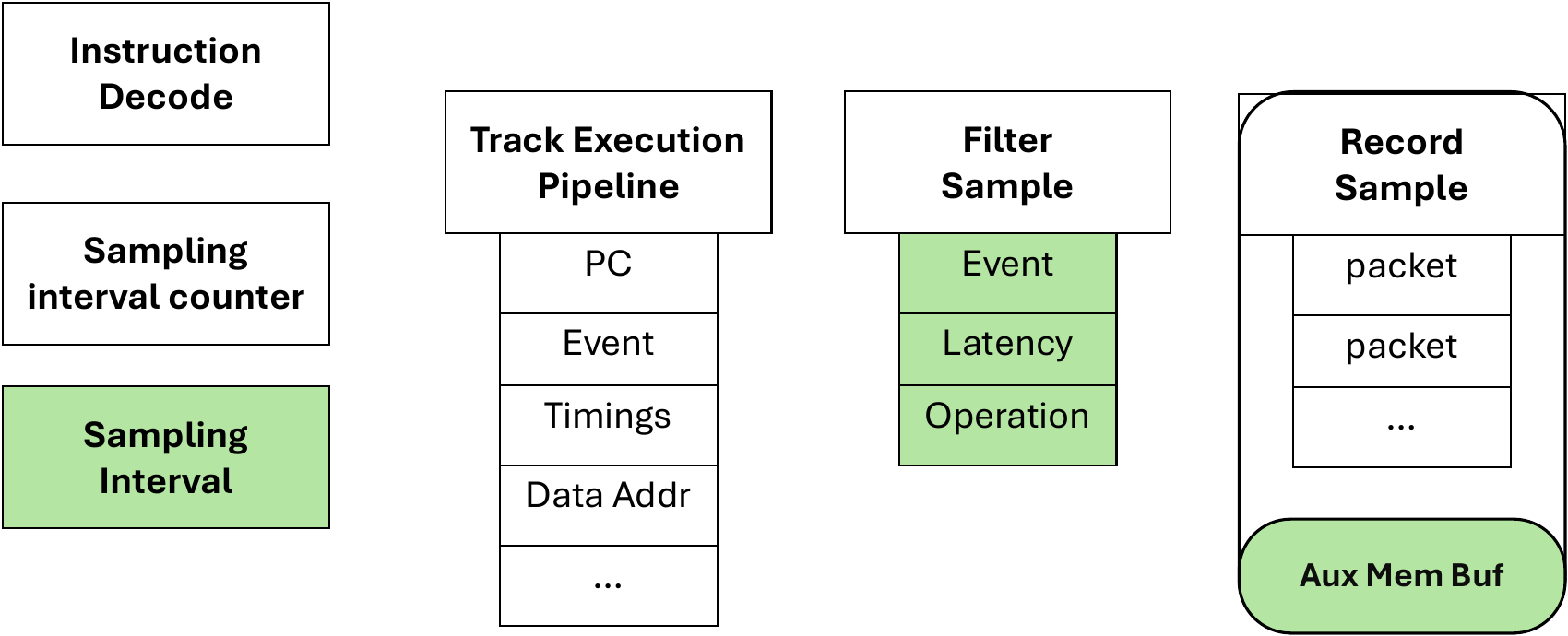}
    \caption{ The main workflow of hardware tracing in ARM SPE. The green blocks can be controlled by the user. }
    \label{fig:arm-spe}
\end{figure}

Figure~\ref{fig:arm-spe} illustrates the main stages of using ARM SPE for precise event sampling. When ARM SPE sampling is enabled, the sampling interval counter is reset to a user-defined value (the sampling period) and then decremented after each operation is decoded. When the counter reaches zero, with some random perturbation added to avoid bias, an operation is selected as a sample from the instruction population. Then, the execution pipeline of the sampled operation within the processor pipeline is tracked to collect detailed information such as timings, memory level, data address, event types, etc. When the operation completes its execution pipeline, the collected information forms a sample record.

A set of programmable criteria is used to filter sample records. Filter criteria related to memory-centric profiling include latency, event type, and memory levels. Sample records that do not fulfill the chosen criteria are discarded while the remaining sample records are saved to a memory buffer, called aux buffer, in the format of a \textit{packet}. When the aux buffer is full, an interrupt is triggered by the processor, and profiling tools will catch the interrupt to process samples and clean up the aux buffer. Consequently, the size of the aux buffer affects the frequency of interrupts and thus impacts the time overhead of ARM SPE. A profiling tool needs to trade off memory overhead (aux buffer size) and time overhead. The sampling mechanism of SPE makes it possible to effectively adjust the sampling overhead, compared to other precise event sampling disturbed by the number of samples. 

\section{Design}
%\begin{figure}
%\centering
%\includegraphics[width=0.5\linewidth]{figures/design.png}
%\caption{The overall architecture of the three-level memory-centric profiler and the emulation platform.\todo{Update to horizontal layout}}
%\label{fig:profiler}
%\end{figure}

The design of the \tool framework aims to provide tools for researchers and developers to perform advanced memory-centric analysis on new architectures using a simple interface.

The \tool runtime component captures memory-related profiling data such as bandwidth utilization and virtual address samples. To minimize the performance impact and enable high accuracy, the data is captured using architecture-specific hardware performance counters. Although this means that porting is required for new architectures, the flexible code ensures the process is as simple as possible. For instance, to collect address samples, the runtime uses SPE when compiling for ARM and PEBS for Intel.

Resource provisioning becomes increasingly dynamic with the adoption of cloud computing practices as well as the emergence of resource disaggregation. Thus, users can benefit from a temporal perspective on profiling results -- different resources may be required in different phases of an application. The \tool design takes this into account by providing a temporal view of key metrics such as memory capacity and bandwidth usage. For instance, an application may have a large memory need during an initialization phase, but a much smaller footprint during the execution phase. A user could take advantage of this by reducing the memory allocated to such a job after initialization is completed.

To enable quick and easy usage, the \tool runtime works in an application-transparent manner and does not require recompiling or changing application-level code.
However, a source-code annotation API enables detailed analysis on a per-kernel and per-object level when necessary. To provide maximum portability across different application programming languages, a simple C API used.

Once profiling data has been captured, flexible post-processing and visualization are enabled by \tool's extensible scripting component. Alongside the provided scripts, users can write their own in Python to process the performance data, enabling advanced application-specific analysis.
%Figure~\ref{fig:design} presents the experimental workflow, where each level of execution collects profiling information that can be visualized separately

\subsection{Memory-Centric Profiling Workflow}

In this paper, we consider a three-stage application profiling workflow enabled by \tool, stretching from simple to advanced metrics.

\textbf{Temporal Capacity Usage.} Memory capacity usage is a fundamental metric. It is crucial to find the right node size for an application and prevent over-provisioning. Viewing the capacity usage over time identifies different resource requirements of separate phases.

\textbf{Temporal Bandwidth Usage.} Memory bandwidth usage is another fundamental metric. For instance, the bandwidth utilization can be used to determine if the application may benefit from HBM and if applications can be colocated on the same node. To track the balance between compute and memory utilization, arithmetic intensity can be profiled by augmenting bandwidth usage metrics with floating-point events. Arithmetic intensity is used to determine if a workload is compute-bound or memory-bound in the Roofline model~\cite{williams2009roofline}. As with the capacity usage, the bandwidth usage may vary in different execution phases of an application, leading to different requirements.

\textbf{Memory Region Profiling.} Region-based profiling enables more advanced profiling use cases in \tool. The virtual addresses of sampled memory accesses are stored and may be used to analyze memory access patterns. For instance, which memory objects are the most accessed inside a certain function? Which objects are seldom read throughout the whole execution?

\begin{comment}
\textbf{Program Level}
The first level of profiling aims to understand an application's requirements on the memory subsystem. These include its arithmetic intensity, memory capacity usage, bandwidth usage, and access pattern. The arithmetic intensity is measured using hardware performance counters, enabling us to place the application into a roofline model. The number of bytes loaded is measured with the \verb|OFFCORE_RESPONSE:L3_MISS| events. The offcore events include all memory loads, including hardware prefetchers. The memory capacity usage is measured by sampling the \verb|numa_maps| file in \verb|procfs|. The memory access pattern is measured in two ways, first using precise event-based sampling (PEBS) to record the virtual address of demand load misses. Secondly, counters related to hardware prefetching are measured to understand if the access pattern is predictable.

\textbf{Memory Usage.}
In a multi-tier environment, we define two key metrics. The \textit{remote capacity ratio} is the ratio of lower-tier memory to total available memory. In our setup, it can be measured from \verb|numa_maps|. The \textit{remote access ratio} is the ratio of memory accesses to a lower tier. In our setup, it is measured using the \verb|LOCAL_DRAM| and \verb|REMOTE_DRAM| offcore events. The event-based sampling of memory accesses is also extended to multiple tiers by separating cache miss events to local or remote memory.

\textbf{Memory Centric Sampling}
\end{comment}

\subsection{Architecture-agnostic Annotation Interfaces}
\tool provides simple high-level C and C++ interfaces for source-code annotations. Its interfaces can be used to create tagged regions that capture either memory allocation or execution phases. An application can use start and end annotations to select a specific code region in execution for profiling, in addition to profiling the whole program execution by default. In the following code snippet, we demonstrate the two types of annotations within \tool:

\begin{lstlisting}[label=lst:api,basicstyle=\ttfamily\footnotesize,frame=single,language=C, 
framexleftmargin=0pt,caption={\tool user-level API examples.}]
nmo_tag_addr("data_a", addr0_start, addr0_end);
nmo_tag_addr("data_b", addr1_start, addr1_end);
...
nmo_start("kernel0");

#pragma omp parallel for
{...}

nmo_stop();
\end{lstlisting}

These high-level annotation interfaces are architecture agnostic so that the same annotated execution phases can be profiled on both x86 and ARM architectures. Internally, \tool changes the architecture-specific settings, including specific hardware events used for profiling memory bandwidth and sampling memory accesses, memory buffers allocation and mapping. In this work, we only focus on the ARM backend in \tool.

\section{Implementation}
We implement the ARM SPE based memory profiling on the top of \tool~\cite{wahlgren2023quantitative}\footnote{https://github.com/KTH-ScaLab/nmo}. \tool is written in standard C++ and utilizes Perfmon2 to read hardware counters and perform sampling. It has been used for Intel's PEBS based sampling in previous works~\cite{wahlgren2022evaluating,wahlgren2023quantitative}. \tool uses the OpenSSL library for generating simple MD5 hashes for the sample trace.

\subsection{ARM SPE for Memory Accesses}
%\todo{Samuel: for this subsection, write as much text as you could following this outline below}

%\todo{Samuel: First, briefly describe the workflow from perf event open, configure perf attributes to select event type, additional setup that is needed for ARM SPE, ring buf and aux buf}

%\todo{Samuel: describe the periodic monitor and get interrupt to process samples: polling which buffer? how to decode packet, filter out packets, etc.}

%\todo{Samuel: explain how other unusual scenarios may occur, such as sample collisions, throttling.}

ARM SPE sampling is enabled in \tool by issuing a \verb|perf_event_open| syscall, which returns a special file descriptor. Two attributes -- the \verb|type| field and the \verb|config| field, are configured and passed to the syscall. The \verb|type| field is set to the hex value \verb|0x2c|, corresponding to the PMU type for ARM SPE. The \verb|config| field is used to set other options passed to ARM SPE. In the current implementation of \tool, this is used to select either a sampling mode in which load instructions, store instructions, or both are sampled. This configuration process is done on a per-core basis. The current implementation of \tool excludes branch instructions in sampling, which can also be sampled with SPE but are known to have sampling bias~\cite{ArmForge} issues on Neoverse N1 hardware. 

\tool uses bitmask to generate the event filter so that only memory access instructions are collected as shown in the third stage in Figure~\ref{fig:arm-spe}. For instance, \verb|0x600000001| corresponds to sampling all loads and stores, consisting of the bits of \verb|2| and \verb|4| mapping load and store sampling respectively. The complexity of configuring the events are transparently within \tool and hidden from the user.

\tool supports user configuration of the selected events for sampling. For instance, if load and store events are selected for ARM SPE, the correct parameters for the attribute struct would be loaded in \tool. Additionally, the field \verb|sample_period| is set to the desired sampling period as configured by the user from the environment variable \verb|`NMO_PERIOD`|. 

%Different from x86 precise sampling in \verb|perf|, in the case of ARM SPE, the processor writes samples to an auxiliary ring buffer, called \textit{Aux Buffer}. 
One difference from x86 precise sampling in \texttt{perf} comes from the extra buffer usage, i.e., aux buffer for ARM SPE. In the buffer mechanism of x86 precise sampling in \texttt{perf}, a \textit{Ring Buffer} is allocated for storing samples in the \texttt{perf\_event} format. When the ring buffer is full, a signal is triggered to call user-defined functions to process samples and clear the buffer. However, for ARM SPE, the processor uses the ring buffer only for recording sample's metadata, i.e., the start address and data size of samples in the \textit{Aux Buffer}, while the detailed information of each sample, including latency spent in the execution pipeline, is actually stored in the \textit{Aux Buffer}. Via the parameter \verb|aux_watermark|, the profiler can adapt the number of samples collected in the aux buffer and the frequency of inserting the sample metadata into the ring buffer.

\tool uses \verb|mmap| to allocate the ring buffer of (N+1) pages, where page is the system page size and on ARM processors in this work 64KB pages are used. The first page in the ring buffer is a metadata page of type \verb|perf_event_mmap_page|, followed by N pages that are directly written to by the kernel and read by the profiler in a producer-consumer model. The aux buffer is also allocated with \verb|mmap|, and its size can be configured by the user.% using the same file descriptor as the regular ring buffer, which corresponds to the value of offset from the start of the ring buffer to the start of the aux buffer, and the size of aux buffer. %The \verb|mmap| for the aux buffer uses the same file descriptor as the regular ring buffer, where the size should be equal to \verb|aux_size| and the offset should be equal to \verb|aux_offset|. Note that the offset is a suggestion to \verb|mmap|, and that the returned address could be placed elsewhere, so it is necessary to calculate the actual offset and update the \verb|aux_offset| field after the \verb|mmap| call is completed for the aux buffer.

During profiling, \tool uses \verb|epoll| to monitor incoming updates to the ring buffer. When ARM SPE generates samples to the aux buffer, \verb|PERF_RECORD_AUX| records the sample metadata, including \verb|aux_offset|, \verb|aux_size|, and \verb|flags| fields, which are saved to the ring buffer. The field \verb|aux_offset| contains the offset in the aux buffer for retrieving an ARM SPE sample. And \verb|aux_size| contains the data size of the sample. Note that ARM SPE records the whole execution pipeline of each sample, and depending on the execution path, such as hit in L1 cache or miss in LLC, each sample may go through different stages in the execution and thus result in different sample sizes. For instance, if a load instruction misses all caches, it will have more recorded data than a load instruction that hits L1. Finally, \tool checks \verb|flags| for the sample status, which could indicate whether data in the buffer has been truncated or if there are any sample collisions present. 

When using ARM SPE for sampling load/store instructions, \verb|perf| records samples in the form of \verb|packets|, as shown in Figure~\ref{fig:arm-spe}, which are 64 bytes large and aligned. For memory access profiling, time stamps and virtual address samples are retrieved by \tool from the packets, where the virtual address is stored as a 64-bit value at an offset of 31 bytes from the base of the packet, and the timestamp is stored as a 64-bit value at the end of the packet at a 56-byte offset from the base. For decoding a packet, the virtual address is prefaced with the byte \verb|0xb2|, and timestamp with the byte \verb|0x71|. Since the trace can contain large gaps between packets, in the implementation of \tool, a packet is skipped from processing if either of those bytes is incorrect, or if the timestamp or virtual address is 0. A invalid packet could be  caused by sample collision if it were sampled before the previous sampled operation has not finished its execution pipeline.

The timestamp timer from ARM SPE uses a different timescale than \verb|perf|, so to maintain API compatibility between different architectures from the perspective of the post-processing scripts, \tool also performs a timescale conversion using the \verb|time_zero|, \verb|time_shift| and \verb|time_mult| fields from the ring buffer metadata page.

\subsection{User-defined Configurations}
\tool supports transparent profiling without the need to modify source codes. This run-time instrumentation relies on preloading the NMO library using the dynamic linker. For this option, \tool also supports a list of environment variables for adapting the configuration based on application's characteristics. For instance, user can adapt the buffer size for ring buffer and aux buffer, which could affect the profiling overhead as shown in the evaluation section. Table~\ref{tab:env} lists the supported list of environment variables. 

\begin{table}[h]
\centering
\caption{A summary of supported Environment Variables and their default values in \tool for user to configure the profiler.} 
\begin{tabular}{|c|c|c|}
\hline
\textbf{Option}    & \textbf{Description}    & \textbf{Default}\\\hline
\verb|`NMO_ENABLE`|  & Enable profile collection  & off \\\hline
\verb|`NMO_NAME`|  & Base name of output files  & "nmo" \\\hline
\verb|`NMO_MODE`|  & Profile collection mode   & none \\\hline
\verb|`NMO_PERIOD`|  & Sampling period  & 0 \\\hline
\verb|`NMO_TRACK_RSS`|  & Capture working set size  & off \\\hline
\verb|`NMO_BUFSIZE`|  & Ring buffer size [MiB]	&1 \\\hline
\verb|`NMO_AUXBUFSIZE`| & Aux buffer size [MiB]&	1 \\\hline 
\end{tabular}
\label{tab:env}
\end{table}

\section{Experimental Setup}

In this work, we use a testbed of ARM Ampere~\cite{ampere} for experiments. The testbed features Arm Neoverse V1 Core~\cite{armneoversev1}. Table~\ref{table:hw} summarizes the hardware specification of the ARM testbed. 

\begin{table}[bt]
\centering
\caption{Hardware specification of the ARM platform used for the experiments.} 
\begin{tabular}{|c|c|}
\hline
\textbf{CPU}    & ARM Ampere® Altra® Max 64-Bit    \\\hline
\textbf{Cores}  & 128 Armv8.2+ cores   \\\hline
\textbf{Frequency}  & 3.0 Hz   \\\hline
\textbf{Mem. capacity}  &  256 GB  \\\hline
\textbf{Mem. technology}  & DDR4   \\\hline
\textbf{Peak bandwidth}  & 200 GB/s   \\\hline
\textbf{L1i}    &  64 KB per core  \\\hline
\textbf{L1d}    &  64 KB per core \\\hline
\textbf{L2}     &  1 MB  per core   \\\hline
\textbf{System Level Cache}  & 16 MB \\\hline
\end{tabular}
\label{table:hw}
\end{table}

The testbed runs Ubuntu 22.04 with Linux kernel 5.15. Docker v27.0 is used for running CloudSuite applications. When applicable, applications were built using GCC 11.4, OpenMPI v4.1.2, and OpenBLAS 0.3.20. In the CloudSuite benchmarks, \tool was bundled into the images. Each test is repeated at least five times, and we report the average value and standard deviation.

We use five benchmarks and applications for evaluation, from Rodinia~\cite{che2009rodinia} and CloudSuite~\cite{ferdman2012clearing} benchmark suites, representing HPC and Cloud workloads. We summarize each application as follows. 

\begin{itemize}
    \item STREAM, implemented in C and OpenMP, is a synthetic benchmark that measures sustainable memory bandwidth. The Triad kernel is reported in this work.
    \item CFD, implemented in C and OpenMP, is an unstructured grid finite volume solver in the Rodinia suite for the three-dimensional Euler equations for compressible flow. 
    \item BFS, implemented in C and OpenMP in the Rodinia suite, performs a breadth-first search on a given graph. 
    \item Page Rank (PR), implemented in Java and Hadoop, is a Graph Analytic benchmark that measures the influence of each vertex on every other vertex.
    %\item Triangle Counting (TR) implemented in Java and Hadoop, is a Graph Analytic benchmark that measures community and connectedness.
    \item In-memory Analytics, implemented in Java and Hadoop, runs the alternating least squares algorithm in memory on a dataset of user-movie ratings.
\end{itemize}
\section{Memory-Centric Profiling Results}

\subsection{Temporal Capacity Usage}

\begin{figure}[ht]
    \centering
    \includegraphics[width=\linewidth]{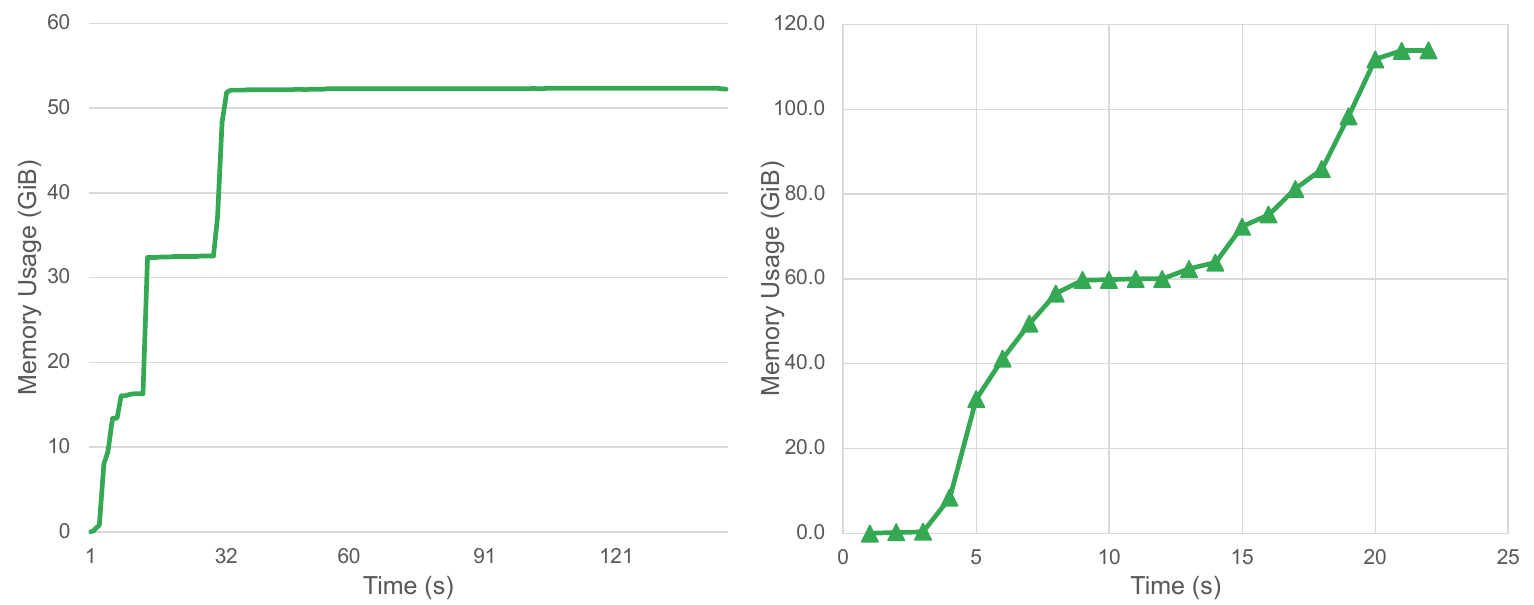}
    \caption{Memory capacity usage over time in Graph Analytics (Page Rank) (right) and In-memory Analytics (left) in CloudSuite.}
    \label{fig:memory-cap}
\end{figure}

%Bandwidth sampling over application execution time.
NMO tracks the total memory capacity usage of target applications, as demonstrated in Figure~\ref{fig:memory-cap}, to guide users in managing and optimizing memory-intensive tasks without storage limitations which may result in unnecessary frequent data swapping. We gave 2 benchmarks in CloudSuite running in a Docker container with 32 cores and 8 GiB memory per core as examples. Memory usage saturates at 52.3 GiB in In-memory Analytics and 123.8 GiB in Page Rank algorithm of Graph Analytics over execution time. The peak memory utilization of these two examples is 20.4\% and 48.4\% respectively, which indicates that less memory resource needs to be reserved for the job, especially for the In-memory Analytics benchmark, to reduce the waste of unused memory. The gradual increase in memory usage also identifies the static allocation of large variables. 

\begin{figure}[ht]
    \centering
    \includegraphics[width=\linewidth]{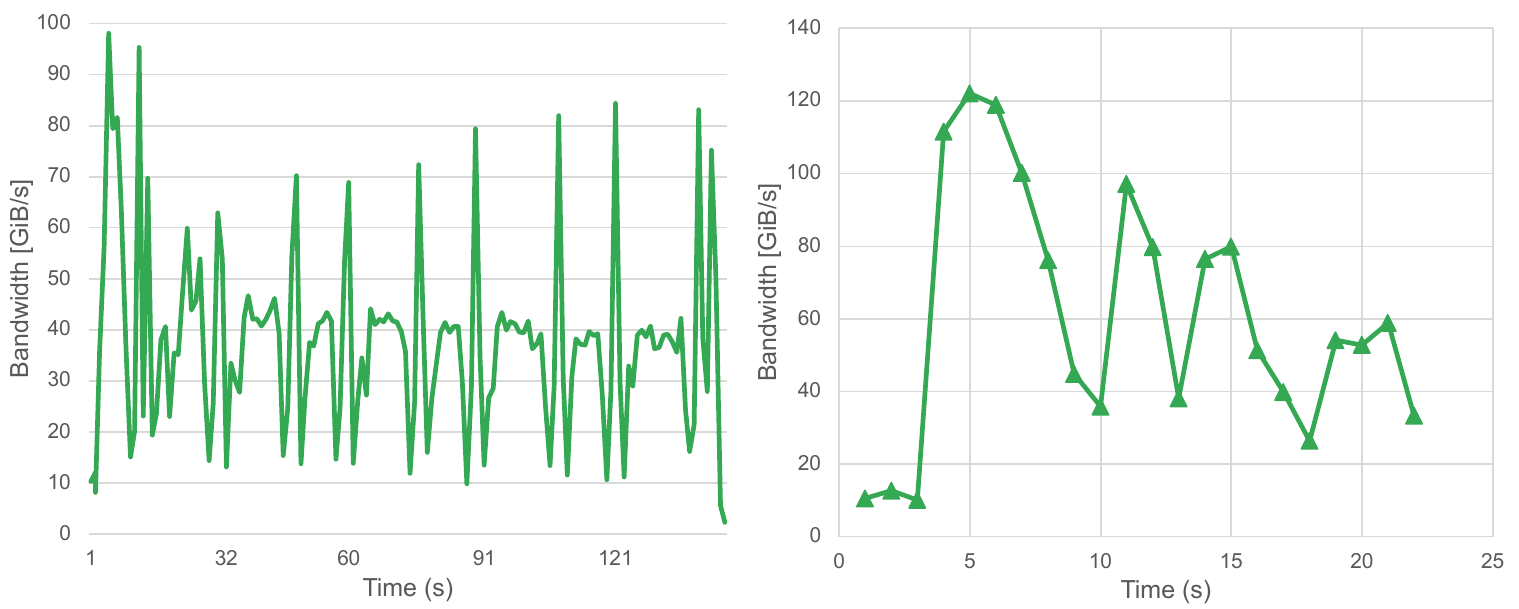}
    \caption{Memory bandwidth usage over time in Graph Analytics (Page Rank) (right) and In-memory Analytics (left) in CloudSuite.}
    \label{fig:bw}
\end{figure}

\subsection{Temporal Bandwidth Usage}

NMO can estimate memory bandwidth based on counting the event of the load and store access on the bus every second, and then dividing the event counter with the length of the interval. It gives users the speed of data access in the target application to improve the bottleneck caused by peak bandwidth requirements of certain processes such as the latency of waiting for data in arithmetic units. Figure~\ref{fig:bw} uses the same examples and demonstrates the memory bandwidth over application execution time. The memory bandwidth usage of In-memory Analytics shows periodic changes about every 15 seconds with nearly 100 GiB/s of peak. This suggests that In-memory Analytics benchmark requires frequent access to in-memory data and its performance is bounded by IO. The bandwidth of Page Rank increases dramatically to 120 GiB/s near 5 seconds, then fluctuates downwards, which potentially demonstrates that the large-scale dataset is loaded at the beginning and the computation steps do not require the fast speed of memory access. 

\subsection{Memory-region Based Profiling}
%Precise Event Sampling using ARM's SPE

\begin{figure}[ht]
    \centering
    \includegraphics[width=0.9\linewidth]{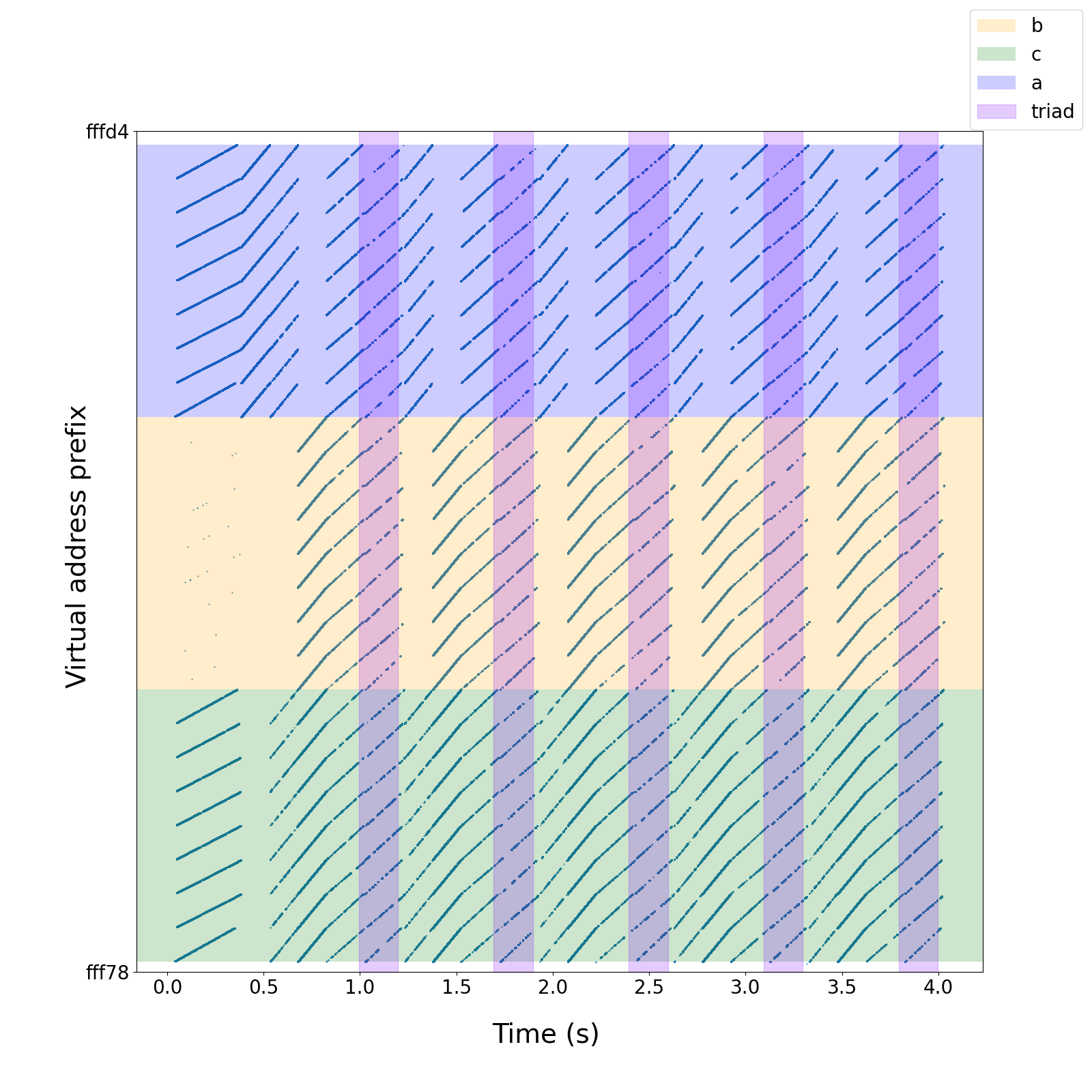}
\caption{Execution phases tagged with sampled memory accesses in the STREAM benchmark on 8 OpenMP threads.}
\label{fig:memtrace-stream}
\end{figure}

\begin{figure}[ht]
    \centering
    \includegraphics[width=0.9\linewidth]{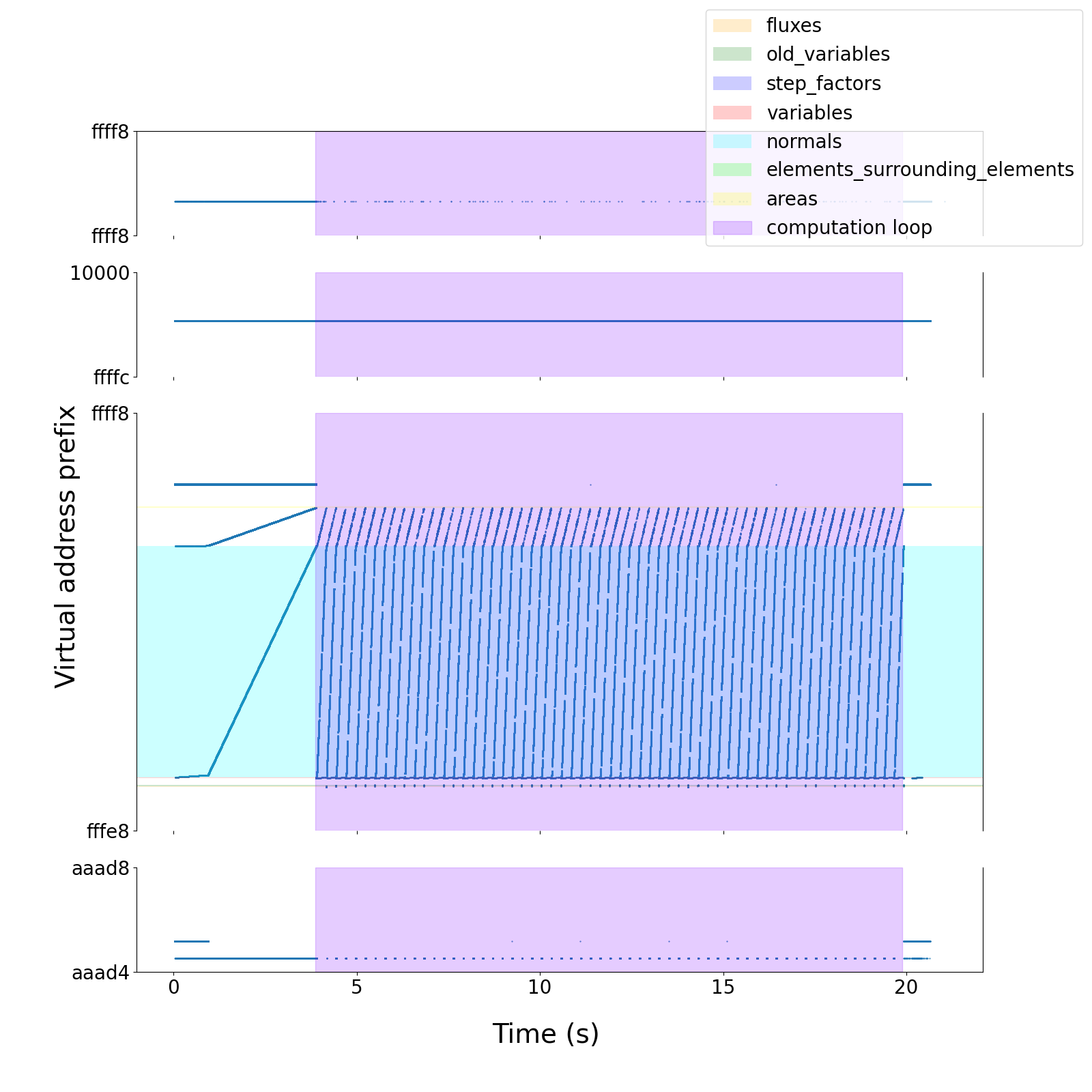}
    \caption{Execution phases tagged with sampled memory accesses in the CFD benchmark at one OpenMP thread.}
    \label{fig:cfd-1thread}
\end{figure}

\begin{figure}[ht]
    \centering
    \includegraphics[width=0.8\linewidth]{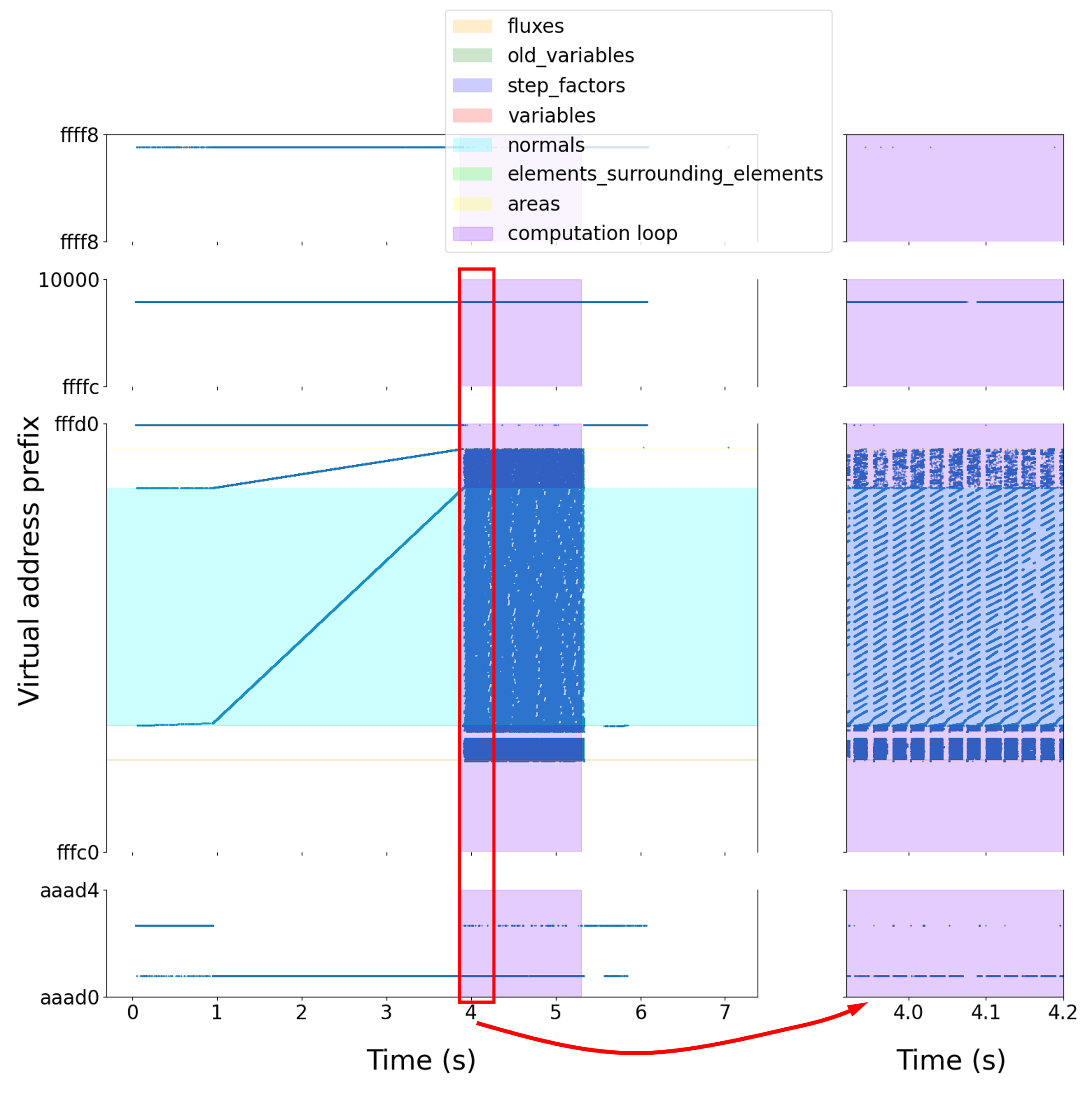}
    \caption{High resolution memory tracing using ARM SPE in NMO in the CFD benchmark at 32 OpenMP threads.}
    \label{fig:cfd-32threads}
\end{figure}

NMO also provides precise event sampling using ARM SPE. The virtual addresses of variables are traced over time in the tagged execution period to analyze the memory access pattern for important variables and memory space. Figure~\ref{fig:memtrace-stream} describes the scatter of sampled memory virtual address in the STREAM benchmarks running 5 iterations with tagged "triad" kernel using 8 OpenMP threads. Blue points are the positions the actual memory load/store taking place. Tags \texttt{a}, \texttt{b}, and \texttt{c} in the legend mark the memory ranges where those variables are stored, and the tags "triad" indicates the execution time of this kernel, computing \texttt{a = b + c * SCALAR} where \texttt{SCALAR} is constant. Executing STREAM benchmark with multithreading, each thread accesses a portion of the array variable in consecutive virtual addresses and each thread accesses roughly the same length, leading to the regular incremental small line segments. Figure~\ref{fig:cfd-1thread} and Figure~\ref{fig:cfd-32threads} compare the memory access pattern of the CFD benchmark at 1 and 32 threads, executing 20 iterations in the ``computation loop'' tag. The memory access at a single thread shows a continuous traverse. However, there were some irregular accesses in the multi-threaded execution and the reduced execution time of the computational kernel in multi-threaded environments obscures their visibility. NMO supports high-resolution memory tracing shown on the right side of Figure~\ref{fig:cfd-32threads} to give more details for the irregular access in parallelization. Unlike STREAM, only \texttt{normals} variable is split properly with a similar length to access in each thread and the other memory region shows an irregular pattern. This irregularity may cause the speedup of the algorithm using multi-thread to be unexpected because of the long latency of memory access.

\section{A Sensitivity Evaluation of ARM SPE}

In this work, we quantify the accuracy of ARM SPE sampling based on the coverage of samples as compared to the profiling results without sampling. We first run a baseline test on the application with the \verb|perf| tool in the \verb|stat| configuration, counting the \verb|mem_access| event, in order to determine the total amount of loads and stores in the execution. Next, we run a comparative test using \tool with ARM SPE enabled to sample load and store events, and the accuracy is determined with

\begin{equation}
    accuracy = 1 - \left| \frac{mem_{counted} - samples*period}{mem_{counted}} \right| 
\end{equation}
where $mem_{counted}$ is the number of load and store events from the baseline test, $samples$ is the number of processed samples, and $period$ is the sampling period determining the value in the sampling interval counter, i.e. if the sampling period is 10,000 then 1 of 10,000 operations will be sampled. %The absolute value was used, in case the test would produce a value which overshoots the expected value which would otherwise produce over 100\% accuracy. 

To quantify the time overhead of \tool in an application, we first run a baseline test that inserts time measurement from the start of the \verb|main| function and until the end of it, to determine how long the application should take to run without any external instrumentation. Then, we run a comparative test with \tool ARM SPE enabled and measure the execution time. Note that the monitoring process in \tool drains the buffer after the exit of the program, which can cause samples to get processed after the interval where the time overhead is measured. However, as we profiling sufficiently long execution periods, influence from the final buffer drain on timing overhead is minimal.

The number of sample collisions is counted by active \verb|PERF_AUX_FLAG_COLLISION| flags, which indicates the dropped records of the Aux buffer when processing samples after filtering.

\subsection{Impact of Sampling Periods}
\begin{figure*}
     \centering
     \begin{subfigure}{0.325\linewidth}
         \centering
         \includegraphics[width=\textwidth]{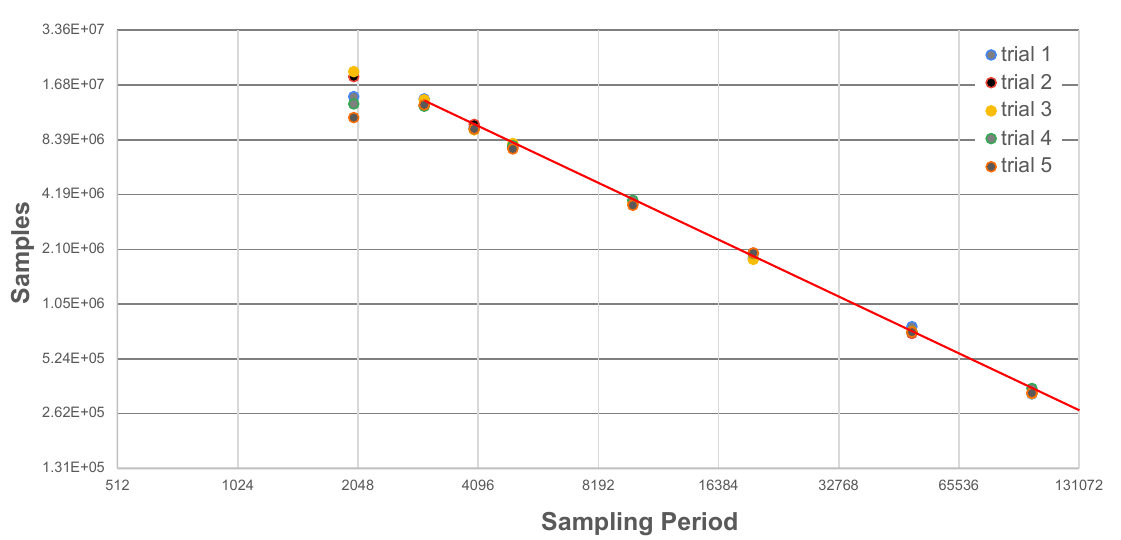}
         \caption{STREAM}
    \label{fig:stream-sample-periods} %Changed name from "mem-time" since it cause name collision
     \end{subfigure}
     \hfill
     \begin{subfigure}{0.325\linewidth}
         \centering
         \includegraphics[width=\textwidth]{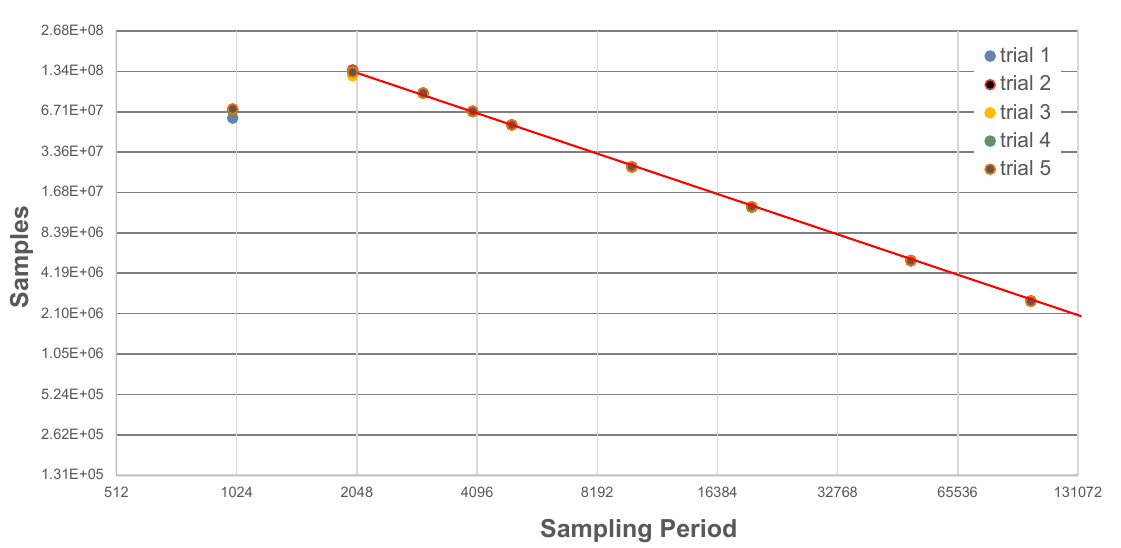}
         \caption{CFD}
         \label{fig:cfd}
     \end{subfigure}
     \hfill
     \begin{subfigure}{0.325\linewidth}
         \centering
         \includegraphics[width=\textwidth]{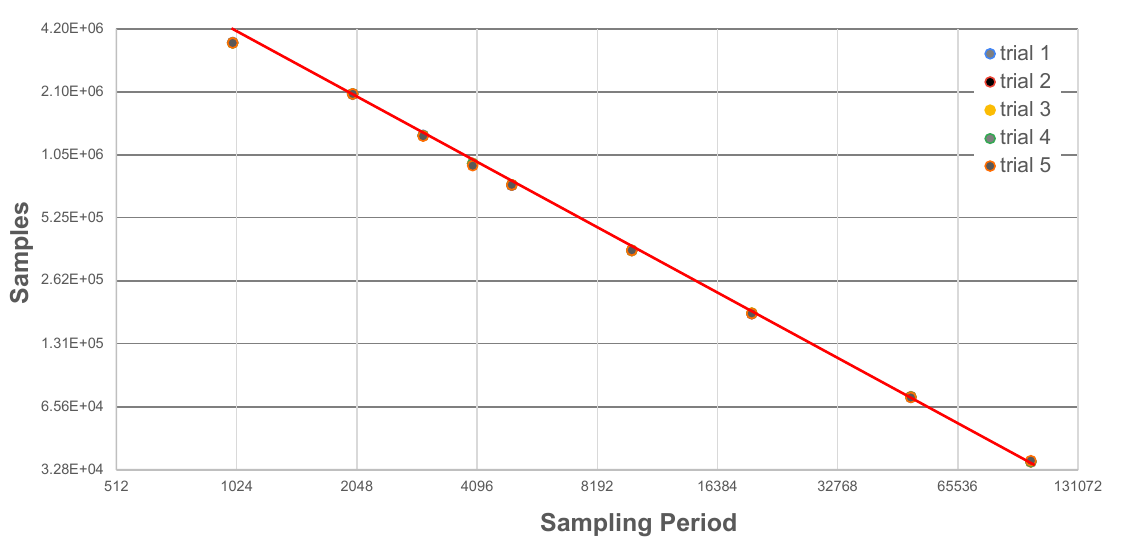}
         \caption{BFS}
         \label{fig:bfs}
     \end{subfigure}
\caption{The number of collected ARM SPE samples of memory accesses in three benchmarks at different sampling periods. At each sampling period, five trials are repeated.}
\label{fig:memtrace}
\end{figure*}

% \todo{Ruimin} First, describe what have been measured and presented here. for each sampling period, five trials are collected. 
% \todo{Ruimin: Observation 1} Linear scaling down, describe quantitatively, on 3 benchmarks. Add explanation: , Takeaway: validate SPE, statistical. 

% \todo{Ruimin: Observation 2} at the smallest sampling period (2000), higher variance in the number of samples being collected. Also, fail to show the linear correlation. For anything abnormal (not as expected), add your explanation / hypothesis. 

\begin{figure*}
     \centering
     \begin{subfigure}{0.325\linewidth}
         \centering
         \includegraphics[width=\textwidth]{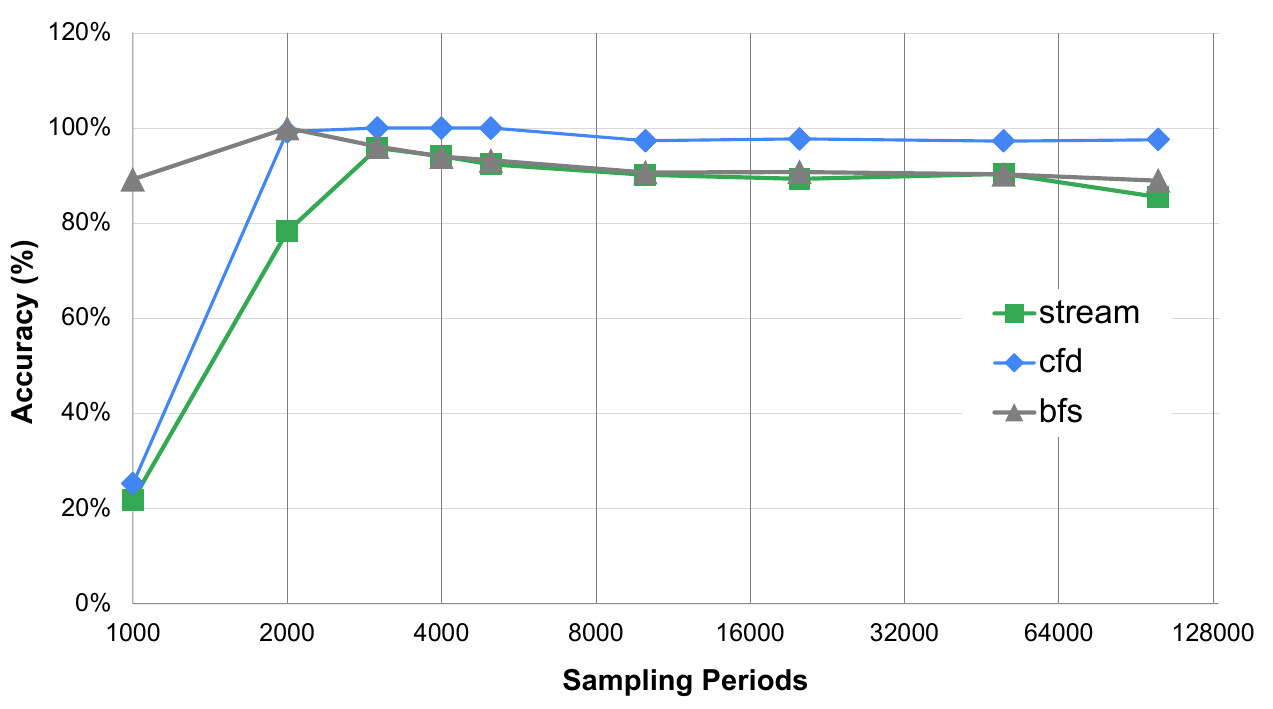}
         \caption{Accuracy}
         \label{fig:accuracy}
     \end{subfigure}
     \hfill
     \begin{subfigure}{0.325\linewidth}
         \centering
         \includegraphics[width=\textwidth]{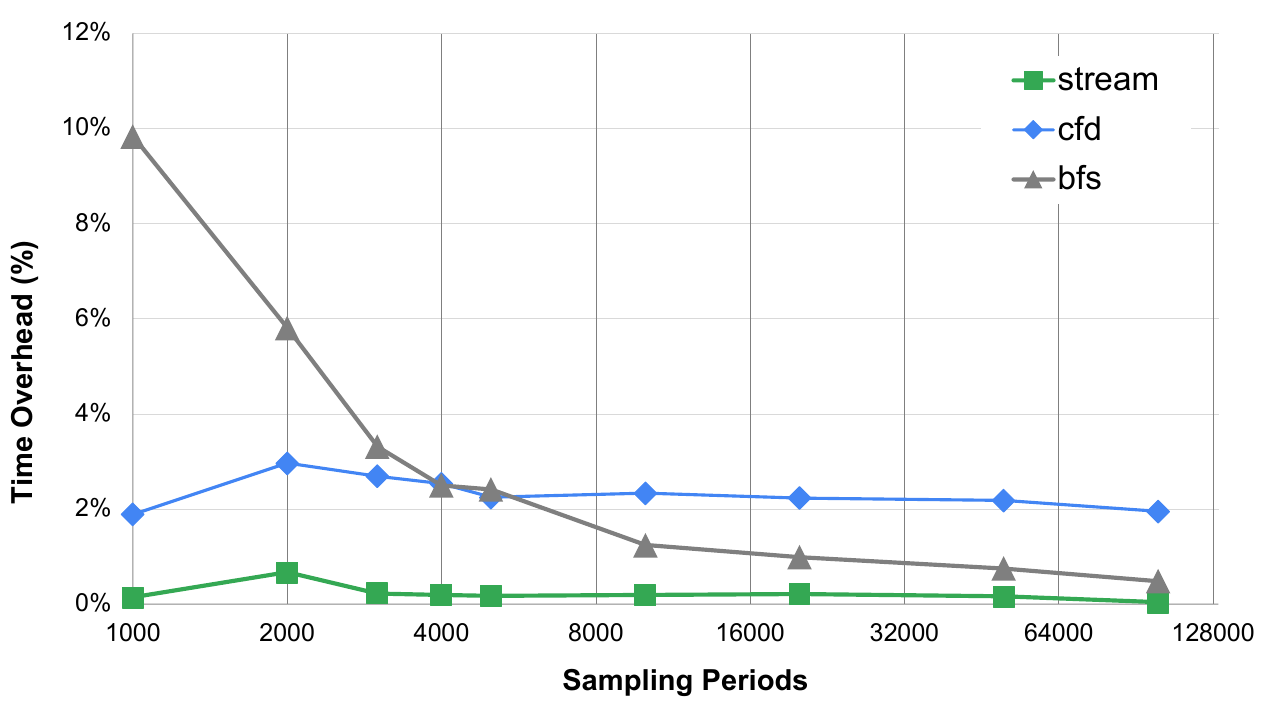}
         \caption{Time Overhead}
         \label{fig:time}
     \end{subfigure}
     \hfill
     \begin{subfigure}{0.325\linewidth}
         \centering
         \includegraphics[width=\textwidth]{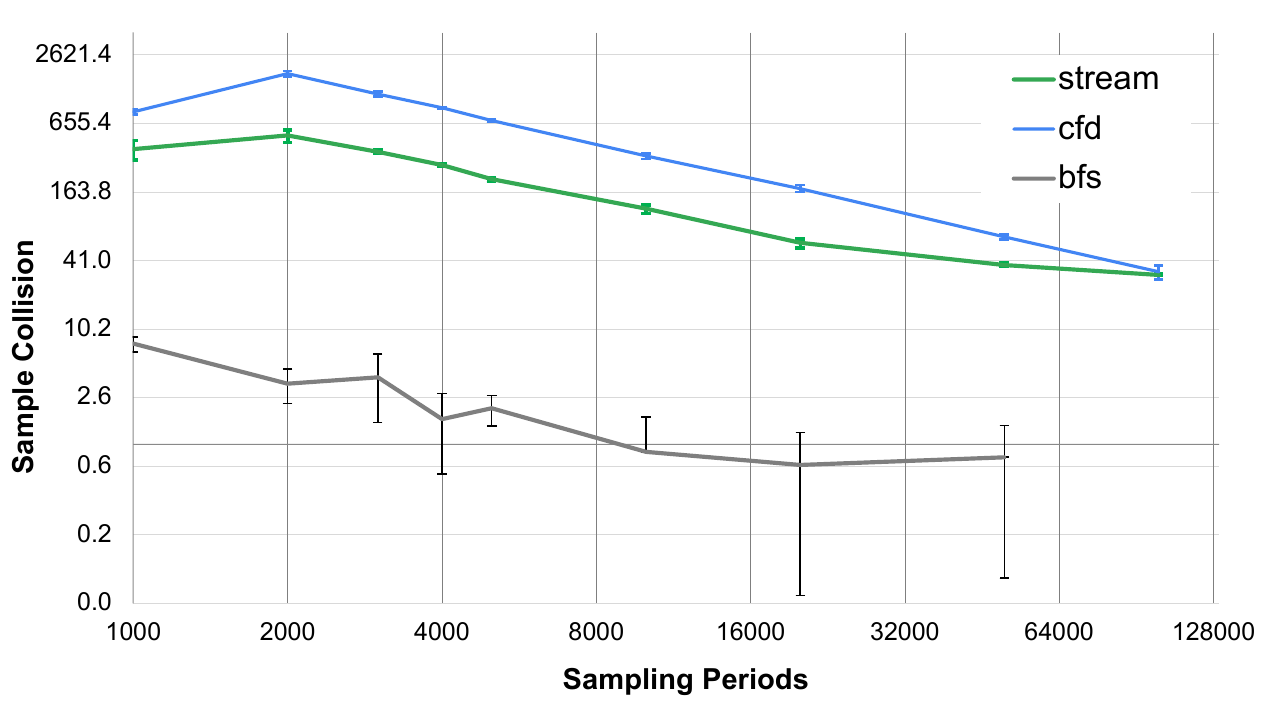}
         \caption{Sample Collision}
         \label{fig:collision}
     \end{subfigure}
\caption{The measured accuracy, time overhead, and sample collision of \tool precise sampling using ARM SPE on memory accesses in three benchmark under different sampling periods.}
\label{fig:sensitivity-periods}
\end{figure*}

Figure~\ref{fig:memtrace} reports the number of collected ARM SPE samples of memory accesses in STREAM, CFD, and BFS benchmarks at different sampling periods. Five trials are collected for each sample period to ensure the generality and identify anomalies. Generally, the number of samples in all applications shows linear scaling down at increasing sampling periods. %, with slopes of about 134, 925, and 12 respectively. (meaningless for this slope?)
This linear regularity validates the sampling process of ARM SPE, which records operations at each sampling interval and indicates that the total number of memory accesses will be obtained by multiplying the sampled memory access with the sampling period. At the smallest sampling period (1000 of STREAM benchmark and 2000 of CFD and BFS), the higher variance in the number of samples being collected is observed. Also, these points fail to show the linear correlation, especially in the CFD benchmark. The abnormal points occur due to the high sampling frequency, which leads to more sample collisions as SPE receives the next sampling command before it has finished processing the previous one, as depicted in Figure~\ref{fig:collision}. These sample collisions happen randomly and uncontrollably, causing fluctuations in the number of samples at the smallest sampling period. 

%\todo{Takeaway: what advice to new users? }

% \todo{Ruimin} First, describe what have been measured and presented here. write as much text as possible. 

% \todo{Ruimin: Observation 1} Linear scaling down, describe quantitatively, on 3 benchmarks. Add explanation: , Takeaway: validate SPE, statistical. 

% \todo{Ruimin: Observation 2} at the smallest sampling period (2000), higher variance in the number of samples being collected. Also, fail to show the linear correlation. For anything abnormal (not as expected), add your explanation / hypothesis. 

 Figure~\ref{fig:accuracy} reports the measured average accuracy of NMO precise sampling using ARM SPE on memory accesses in STREAM, CFD, and BFS benchmarks at different sampling periods. The accuracy of both STREAM and CFD benchmarks shows a sharp increase in the sampling period below 3000 and then tends to stabilize at 94-96\% with a slight decrease. The accuracy of the BFS benchmark is prominently higher than the other two at smaller sampling periods. As shown in Figure~\ref{fig:collision}, the sample collision reaches up to even 510 and 1780 in STREAM and CFD respectively while that keeps below 10 in BFS although all of them tend to decrease with the sampling period. This shows that sample collision caused by inadequate sample interval for processing the last sample is one of the important factors affecting the accuracy of memory access profiling. However, the trend of increasing accuracy and decreasing sample collision is not always consistent. The slight decrease in accuracy may result from fewer captured samples and loss of changing information at sampling intervals to estimate an accurate memory access model. 

 Figure~\ref{fig:time} reports the time overhead at different sampling periods. Time overhead is the proportion of the total execution time that is spent on executing NMO using ARM SPE. As the sampling period decreases below 4000 in the BFS benchmark the time overhead increases with a higher slope while a non-significant decline occurs contrary to that in STREAM and CFD. BFS has the largest time overhead at sampling periods below 4000 because it has the highest amount of samples, shown in Figure~\ref{fig:bfs}. When a sample is recorded in ARM SPE, it is written to the memory buffer after a filter. The main time overhead comes from processing samples after the interrupt from SPE when the buffer is full. A smaller sample period means a higher sampling frequency which leads to more samples being written to the memory buffer, and the time overhead increases as more frequent interrupt requests occur. When sample collision before filtering happens, the samples are directly discarded without being filtered and written to the buffer, which means no time overhead. So the low time overhead of STREAM and CFD is dominated by the number of discarded samples when the sampling period is 1000 or 2000. In conclusion, users are supposed to avoid using a small sampling period below 2000 for high accuracy and less variation. Considering time overhead, 10,000 to 50,000 are suggested.

%\todo{Takeaway: what advice to new users? }

\subsection{Impact of Aux Buffer}
\begin{figure}
    \centering
    \includegraphics[width=\linewidth]{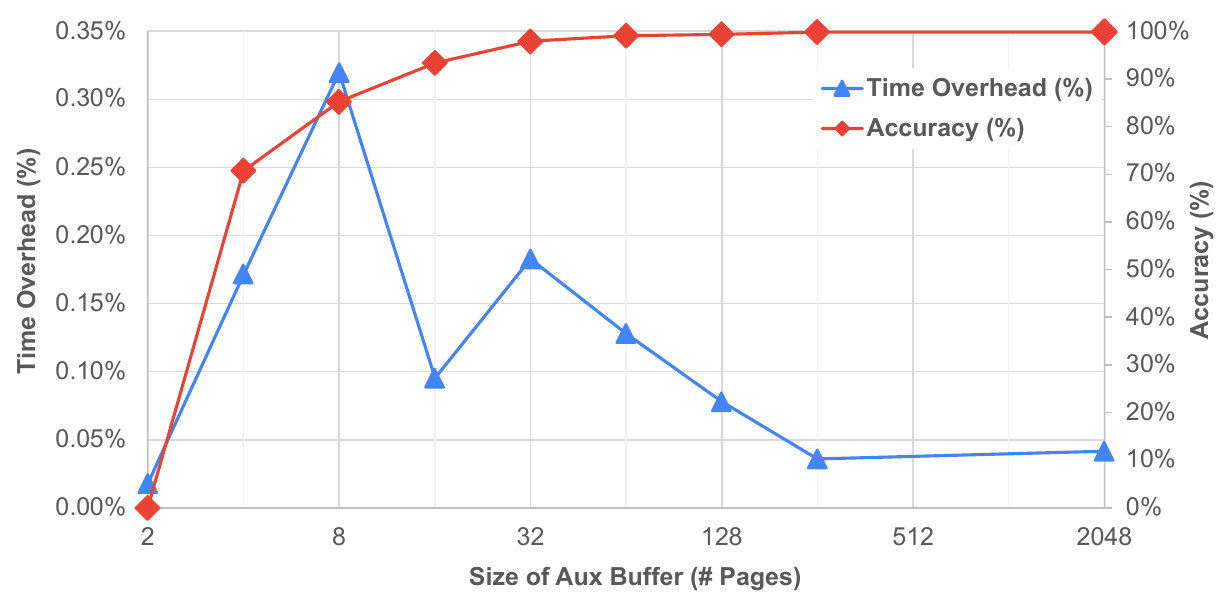}
    \caption{The impact of the Aux buffer size in ARM SPE profiling on time overhead and accuracy on the STREAM benchmark.}
    \label{fig:mem-time}
\end{figure}

 Figure~\ref{fig:mem-time} presents the changes in time overhead and accuracy of NMO with different Aux buffer sizes in ARM SPE profiling, tracing triad application in STREAM benchmark with 32 threads and 1G array size. The size of the ring buffer is fixed to 9 pages.
 
\subsubsection{Time Overhead}
% Observation 1: lowest time overhead at the smallest buffer size, why?

% Observation 2: beyond 0.5 GiB, a larger buffer size reduces the time overhead, why? interrupt the program execution fewer times to process samples. 

The lowest time overhead occurs at the smallest buffer size and then rapidly increases when going from 2 to 8 pages which indicates that ARM SPE loses all samples if the Aux buffer is not large enough. The minimum size to ensure SPE works is 4 pages. Beyond 32 pages, a larger buffer size reduces the time overhead due to the fewer times to interrupt the program execution to process samples.

\subsubsection{Accuracy}
% Observation 1: more mem buf, higher accuracy. refer to the figure, quantitative / specific, sustain at 0.99\% accuracy. Add your explanation: 

% Insights (): beyond xxx GiB, furthering increasing aux buf size bring minimal benefits to accuracy. 

%Accuracy increases steadily with an increased amount of Aux buffer pages because of the improved capability of recording samples of the Aux buffer. Increasing the amount of Aux buffer pages will increase the amount of SPE data that can be processed by MAP, increasing the number of SPE samples obtained. It achieves stability of over 99\% at 64 pages and further spending more memory for Aux buffer will bring minimal benefits to accuracy. 

Accuracy increases steadily with an increased amount of Aux buffer pages. When the Aux buffer is larger, it reduces the amount of time where samples can collide. At over 64 pages, an accuracy of over 99\% is achieved and further increases to the Aux buffer size provide minimal benefits to accuracy.

An Aux buffer size of 16 pages is a good choice to balance the trade-off between the accuracy of profiling memory accesses and the time and memory buffer overhead with about 0.2 GiB memory and 0.1\% time overhead to achieve 93\% accuracy.
 
\subsection{Impact of Threads}
\begin{figure}[ht]
    \centering
    \includegraphics[width=\linewidth]{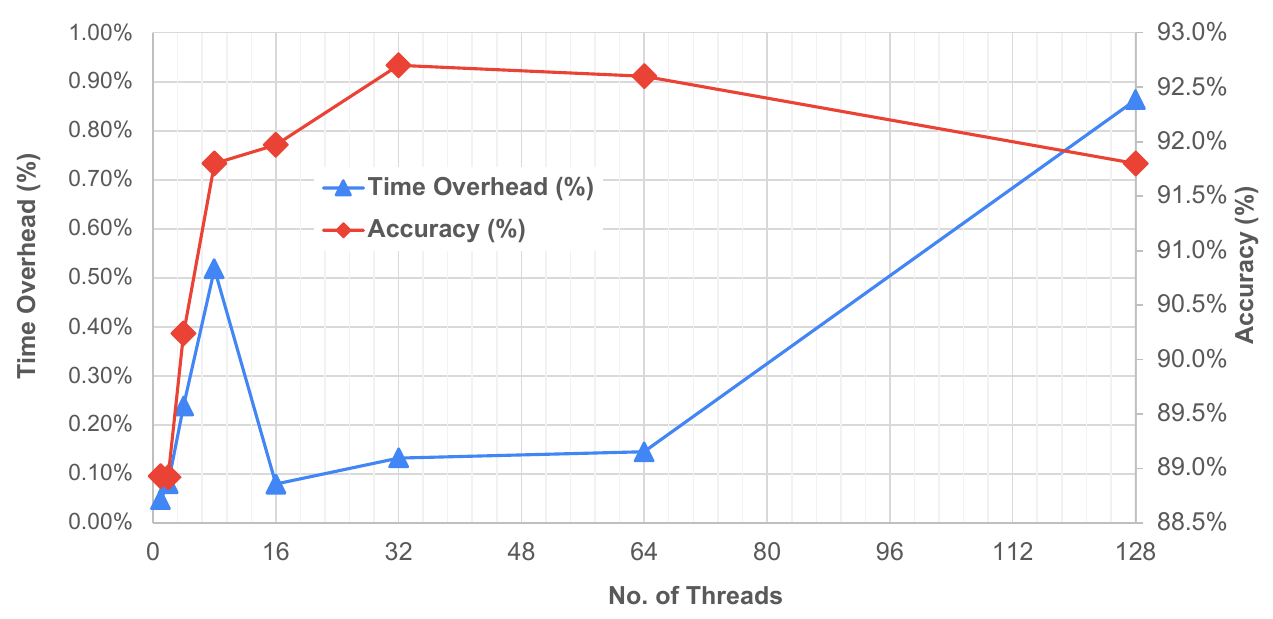}
    \caption{The measured time overhead and accuracy of \tool using ARM SPE on memory accesses from the STREAM benchmark at an increased number of OpenMP threads.}
    \label{fig:accuracy-stream}
\end{figure}
\begin{figure}[ht]
    \centering
    \includegraphics[width=\linewidth]{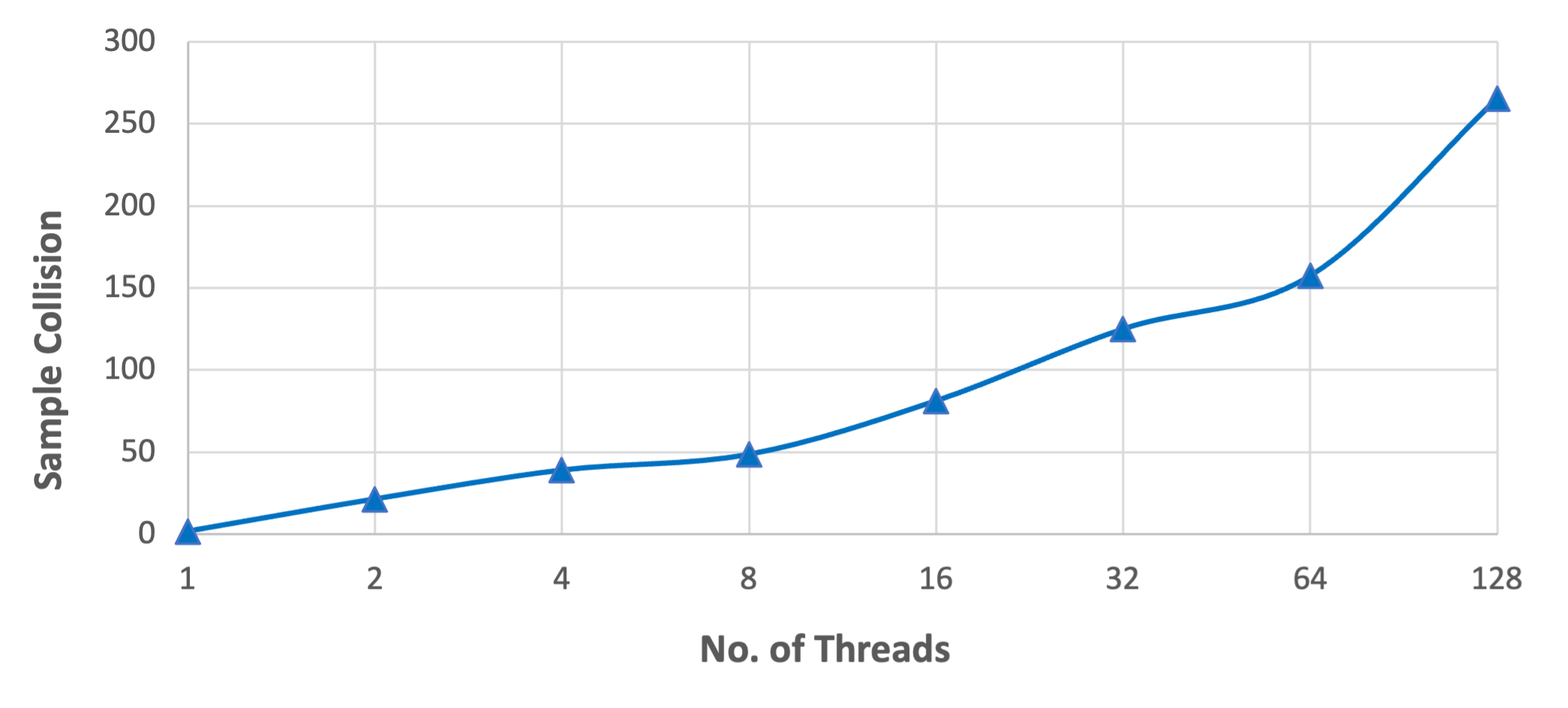}
    \caption{The sample collisions from the STREAM benchmark at an increased number of OpenMP threads.}
    \label{fig:throttle}
\end{figure}

Figure~\ref{fig:accuracy-stream} illustrates the variation of time overhead and accuracy with the increased number of OpenMP threads from the same setup of STREAM benchmark as Figure~\ref{fig:mem-time}, keeping Aux buffer size at 16 pages. Time overhead has a general trend of gradual increase, with the highest 0.86\% time overhead at 128 threads in addition to the number of threads at the two anomalies of 4 and 8 points.
  
The accuracy of NMO is not sensitive to the number of threads on STREAM benchmark and sustains at 89-93\%. It initially increases with the number of threads, reaching the peak around 93\% at 32 threads. One possible explanation is that \tool allocates an Aux buffer for each thread. Thus, by running fewer threads, the same amount of samples is written into fewer buffers, effectively reducing the buffer size. We also observed reduced accuracy at a high thread count in Figure~\ref{fig:accuracy-stream}, by inspecting the sampling collisions in perf at increased OpenMP threads in Figure~\ref{fig:throttle}, we can see a substantial increase in sampling throttling at a high thread count, which could be the cause for reduced samples collected from ARM SPE.

%The lower accuracy at fewer threads is caused by the non-compensation from over-sampled threads. SPE will collect data from the entire program ran across all threads. So the combined statistics for multiple threads can compensate for the lack of sampling in a single thread. This feature is similar to that of AMD IBS as Last Committed Instruction(LCI) sampler~\cite{sasongko2023precise}.

%\todo{I can't find anything about Last Committed Instruction, what is that? One of the explanations we talked about was the fact that Aux buffer allocation is done per thread, so by running fewer threads the same amount of samples is crammed into one buffer instead of spread across many larger buffers, making the effective buffer size smaller. The reason why it becomes lower at higher thread counts, the only theory I can come up with is that the STREAM benchmark peaks in performance at 32 threads and reduces at 64 and 128 threads, and the same performance loss is affecting sample processing. /Samuel}

\section{Related Works}

\textbf{PMU-based profiler.} Mainstream performance monitoring units in modern processors include Intel PEBS, AMD IBS and Arm SPE. Weaver et al.~\cite{weaver2008can,weaver2013non} explores the stability of event counts and the reliability of using hardware performance counters for profiling with 11 implementation methods on nine x86 architectures. The ScaAnalyzer profiler~\cite{liu2015scaanalyzer} is built based on IBS and PEBS to analyze the scalability bottlenecks of non-uniform memory access (NUMA) architectures. Chen et al.~\cite{chen2020atmem} develops a profiler based on Intel's PEBS for adapting the profiling overhead and accuracy at runtime in multi-threaded graph processing applications and migrating hot memory regions into fast memory tiers. Boehme et al.~\cite{boehme2016caliper} develops Caliper, a cross-stack general performance introspection tool for HPC applications. Caliper separates developers and users and generalizes the data model by providing the Caliper annotation API. ComDetective~\cite{sasongko2019comdetective} monitors PMU events and debug registers to detect communication overhead in multi-threaded programs. Mess benchmarks~\cite{esmaili2024mess} reports bandwidth-latency curves for characterizing memory system by hardware counters and pointer chasing. The Mess benchmark is integrated in actual systems and also simulators, including x86, ARM, RISC-V processors. Many memory performance analysis tools utilize PMUs for low-overhead profiling. However, the majority of them focus on Intel's PEBS and AMD's IBS, while few studies look into ARM's SPE because HPC systems are dominated by x86 processors. With the emergence of ARM processors for HPC and Cloud, our study on using ARM SPE for memory analysis fills this gap.

\textbf{Quantitative evaluation of profiler.} Correctness and accuracy of precise event-based sampling profiling methods are an active research area. Sasongko et al.~\cite{sasongko2023precise} quantifies and analyzes the accuracy, stability, sampling bias, overheads, and functionalities of the PEBS and IBS sampling capabilities. Their conclude that PEBS gives more accurate event sampling, while IBS provides more information per sample. Xu et al.~\cite{xu2019can} discovers that the instruction measurement inaccuracy in event-based sampling profilers is caused by skid and propose software techniques to improve the accuracy. Yi et al.~\cite{yi2020precision} studies the sampling bias of PEBS and proposes to artificially insert NOP instructions to improve the precision. Soramichi et al.~\cite{soramichi2017quantitative} studies the overhead of PEBS and quantifies 200--300~ns of CPU overhead per PEBS event at high sampling rates for online system-noise analysis. Gottschall et al.~\cite{gottschall2021tip} proposes the Oracle profiler as a golden reference to reveal the shortcomings of existing profilers that inaccurately attribute performance loss to specific instructions. By combining the attribution policies of Oracle with statistical sampling they propose a practical implementation called the Time-Proportional Instruction Profiler.

\section{Conclusion and Future Works}
In this work, we proposed a memory-centric profiling tool called \tool for ARM processors. Besides memory bandwidth and capacity usage, \tool specifically leverages ARM's Statistical Profiling Extension (SPE) to enable memory-region-based profiling. We evaluated \tool in five applications, including Stream, Rodinia's CFD and BFS, and CloudSuite's Page Rank and In-memory Analytics. On a recent ARM Ampere processor, our results quantitatively evaluate the time overhead and sampling accuracy of ARM SPE at different sampling periods and aux buffer sizes. At 3000 and 4000 sampling periods, the ARM SPE profiling achieves the highest accuracy above $94\%$ at a time overhead of $0.2\%$--$3.3\%$. We found that a sampling frequency lower than 2000 causes significant sample drops and low accuracy. Meanwhile, aux buffer sizes of 16--32~pages result in the optimal overhead and accuracy in the tested applications. For future works, we plan to continue the evaluation of the bias when sampling the same event in different positions of code. Furthermore, we will quantify the impact of bias on memory profiling and provide more advanced metrics, such as tracing cache activities.

\section*{Acknowledgment}
This research is supported by the European Commission under the Horizon project OpenCUBE (101092984). This work was performed under the auspices of the U.S. Department of Energy by Lawrence Livermore National Laboratory under Contract DE-AC52-07NA27344. LLNL-CONF-869756.

\bibliographystyle{IEEEtran}
\bibliography{main}

\end{document}